# Hierarchical network design for nitrogen dioxide measurement in urban environments, part 1: proxy selection

Lena Weissert[1,3], Georgia Miskell[1], Elaine Miles[2], Kyle Alberti[2], Brandon Feenstra[4], Hamesh Patel[3], Vasileios Papapostolou[4], Andrea Polidori[4], Geoff S Henshaw[2], Jennifer A Salmond[3], David E Williams[1,*]

*Email david.williams@auckland.ac.nz ph +64 9 923 9877

1. School of Chemical Sciences and MacDiarmid Institute for Advanced Materials and Nanotechnology, University of Auckland, Private Bag 92019, Auckland 1142, New Zealand

2. Aeroqual Ltd, 460 Rosebank Road, Avondale, Auckland 1026, New Zealand

3. School of Environment, University of Auckland, Private Bag 92019, Auckland 1142, New Zealand

4. South Coast Air Quality Management District, 21865 Copley Drive, Diamond Bar, CA 91765, USA

Abstract

Previous studies have shown that a hierarchical network comprising a number of compliant reference stations and a much larger number of low-cost sensors can deliver reliable air quality data at high temporal and spatial resolution for ozone at neighbourhood scales. Key to this framework is the concept of a ‘proxy’: a reliable (regulatory) data source whose results have sufficient statistical similarity over some period of time to those from any given low-cost measurement site. This enables the low-cost instruments to be calibrated remotely, avoiding the need for costly on-site calibration of dense networks.

This paper assesses the suitability of this method for local air pollutants such as nitrogen dioxide which show large temporal and spatial variability in concentration. The ‘proxy’ technique is evaluated using the data from the network of regulatory air monitoring stations measuring nitrogen dioxide in Southern California to avoid errors introduced by low-cost instrument performance. Proxies chosen based on land use similarity signalled typically less than 0.1% false alarms. Although poor proxy performance was observed when the local geography was unusual (a semi-enclosed valley) in this instance the closest neighbour station proved to be an appropriate alternative. The method also struggled when wind speeds were low and very local sources presumably dominated the concentration patterns. Overall, we demonstrate that the technique can be applied to nitrogen dioxide, and that appropriate proxies can be found even within a spatially sparse network of stations in a region with large spatio-temporal variation in concentration.

Keywords

air quality sensors; nitrogen dioxide; low-cost sensor network; calibration

## *1. Introduction*

In the last several years attempts have been made to supplement regulatory monitoring networks with low-cost instruments to measure air pollutant concentrations at a higher spatial and temporal resolution than previously possible (Bart et al., 2014; Snyder et al., 2013; Weissert et al., 2018). However, the performance of low-cost gas-phase sensors is compromised by issues related to interactions with other gases, as well as meteorological conditions, long-term stability or drift in calibration (Lewis et al., 2016). Consequently, rigorous maintenance and calibration procedures are necessary to ensure long-term data quality and reliability from low-cost sensors, which increases the costs significantly. Common calibration procedures involve factory calibration, where the sensors are calibrated in the laboratory under controlled conditions, and field calibration, where low-cost sensors are co-located with regulatory monitoring instruments. The former approach is often not appropriate due to interactions with other gases as well as the effects of different meteorological conditions (temperature, humidity, wind speed) that are not accounted for (Bigi et al., 2018; Cross et al., 2017; Lewis et al., 2016; Spinelle et al., 2017). The latter approach must be repeated periodically to ensure data reliability, which is not only resource and time consuming, but can also lead to data gaps, unknown errors associated with long-term drift and issues related to sensor handling and transport during relocation (Bigi et al., 2018).

To address these issues, we recently explored simple and effective solutions for implementation within a hierarchical network comprising both low-cost and regulatory-grade instruments that allow remote calibration of sensor data (Miskell et al., 2018) and identification of sensor drift (Miskell et al., 2016). We introduced the idea of using a sparse network of well-maintained regulatory-grade instruments to provide a reliable proxy data set which can be used to verify reliability and calibrate data from low-cost instruments which are deployed in a spatially much denser network (Miskell et al., 2016, 2018). The important concepts: a proxy model, a measurement model and a parameter estimation model; have been described in detail in Miskell et al. (2019).

The ideas have been tested using ozone ($O_3$) data derived from a network of low-cost semiconducting oxide-based devices around the Lower Fraser Valley (LFV), British Columbia, Canada (Miskell et al., 2018), which has a relatively smooth field of $O_3$ concentrations and where it was easy to find a reliable proxy for $O_3$. To test the transferability of this approach to a more complex environment where land use is much more variable, we also successfully tested the use of proxies for $O_3$ correction and calibration in Southern California, USA (Miskell et al., 2019). These studies used basic land use similarity and proximity as criteria for the selection of a suitable proxy dataset (Miskell et al., 2016; Miskell et al., 2018). Ozone concentrations are mostly dependent on sunlight, traffic density and meteorology (wind direction, speed, boundary layer depth). Sites in the LFV were grouped into urban, residential and rural based on the dominant land use within 1 km of the site (Miskell et al., 2018) whereas sites in Southern

California were grouped by proximity and by traffic density (Miskell et al., 2019). While this approach was successful for $O_3$, it still needs to be tested using other common urban air pollutants, such as nitrogen dioxide ($NO_2$), which varies considerably over short distances (Deville Cavellin et al., 2016; Li et al., 2019; Weissert et al., 2019; Weissert et al., 2018).

In the present paper, we address the problem of identification of suitable proxies for a network of $NO_2$ sensors, and to identify conditions under which the proxy selection might fail.

Nitrogen dioxide is generally a secondary pollutant, which is formed by oxidation of nitrogen oxide (NO) mostly related to traffic and industrial sources. A popular method to predict the spatial variability of $NO_2$ concentrations is land use regression (LUR) analysis, where $NO_2$ concentrations are modelled using a set of land use variables (e.g. distance to road, traffic density or land cover), that are typically available through geographic information systems (GIS). Numerous cities, particularly across Europe and North America, have developed LUR models to estimate the spatial distribution of air pollutants. In this study we aim to: 1) identify land use variables that are most commonly reported as being significant predictors for the spatial variability of $NO_2$ concentrations; 2) use these land use variables to identify suitable proxies for $NO_2$ based on land use similarities; and 3) compare the performance of selected proxies based on land use similarities with nearest proxies in terms of location. This study offers some important insights into the possibilities of remotely calibrating low-cost $NO_2$ sensors, which will be essential to effectively manage and maintain large low-cost sensor networks.

## *2. Methods*

### *2.1 Study sites*

The approach is using the regulatory network sites around two regions; Los Angeles (LA) and the Inland Empire (IE) which includes Riverside and San Bernardino Counties (Figure 1, Table 1) in Southern California. All sites are equipped with automated reference method nitrogen oxides (NOx) analyzers, which are maintained and regularly serviced by the South Coast Air Quality Management District (South Coast AQMD). Specifically, eight sites are equipped with a model 42i NOx analyzer by Thermo Fischer Scientific (Franklin, MA), while the Fontana site is equipped with a model 200E NOx analyzer by Teledyne Advanced Pollution Instrumentation (San Diego, CA). The regulatory sites are selected to be representative for locations with high pollutant concentrations, or high population exposure, or source impact or background. We used hourly-averaged data from January – July 2018 collected at 9 sites ($n$ = 5 in LA, $n$ = 4 in IE). Measurements are mixing ratios: parts-per-billion ($10^9$) by volume (ppb).

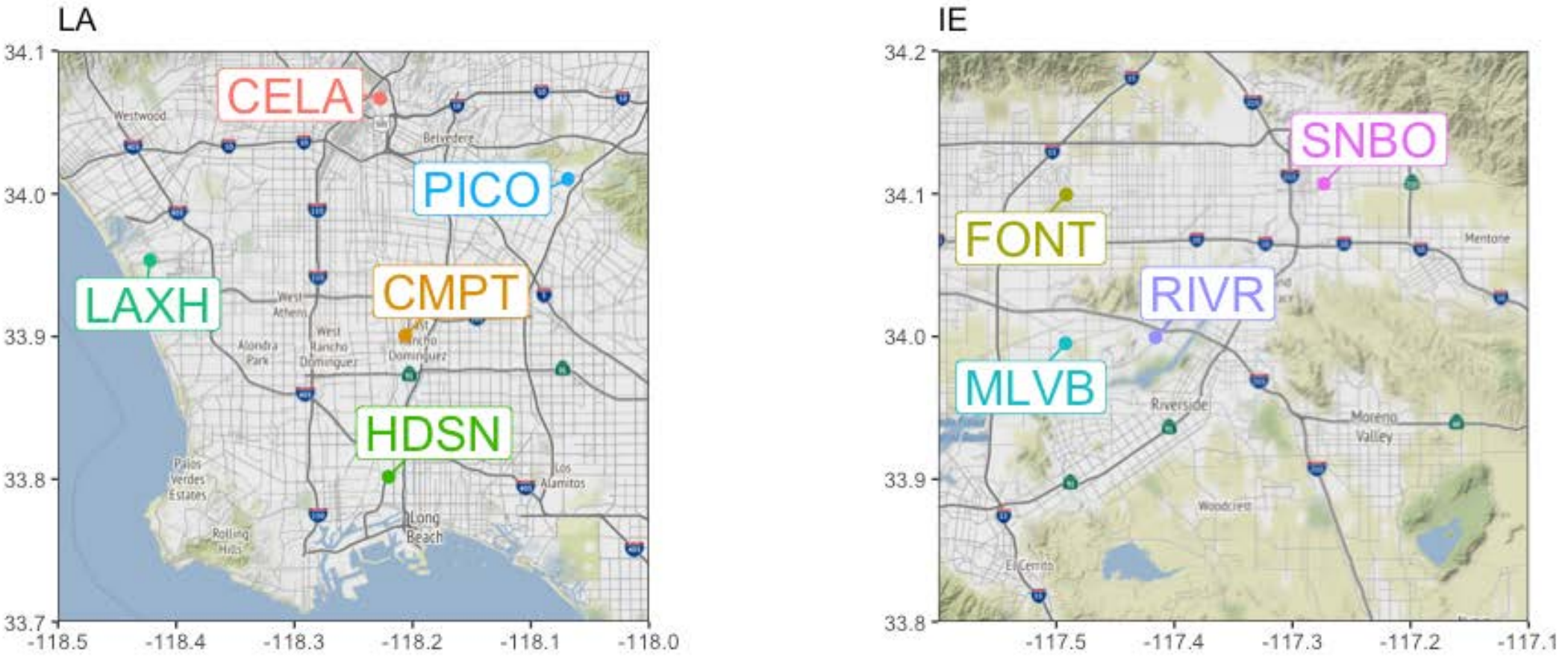

Figure 1. Map of the two regulatory monitoring networks.

Table 1. Descriptions for the nine regulatory locations. Land uses are based on publicly available data.

| **AQS Name** | **AQS ID** | **Dist. to motorway / m** | **Elevation above sea-level / m)** | **Freeway and primary road length < 1 km / m** |
|---|---|---|---|---|
| Rubidoux | RIVR | 685 | 248 | 6708 |
| Mira Loma | MLVB | 2480 | 220 | 0 |
| San Bernardino | SNBO | 2620 | 316 | 2408 |
| Fontana | FONT | 3210 | 363 | 5889 |
| Pico Rivera | PICO | 803 | 58 | 6563 |
| Compton | CMPT | 1660 | 22 | 7040 |
| LAX Hastings | LAXH | 4450 | 37 | 4270 |
| Long Beach (Hudson) | HDSN | 1150 | 10 | 5566 |
| Central LA | CELA | 917 | 89 | 5168 |

*2.2 General characteristics of the data*

The temporal and spatial variability of $NO_2$ concentrations was, as expected, large: Figure 2. The distribution of values varied month-by-month, from broad and bimodal in winter (mean ±sd temperature/RH: 16 ±7ºC/46 ±26%) to narrower and monomodal in summer (mean ±sd temperature/RH: 27 ±6ºC/55 ±21%), and displayed different patterns at different sites (Figure 2a). Diurnal variations were irregular: in winter, the variation was typically small fluctuations upon a large and variable background; in summer, values were frequently low and hardly varying (Figure 2b). The diurnal variation showed patterns that were frequently similar across a number of sites whilst being very different at others (Figure 2c).

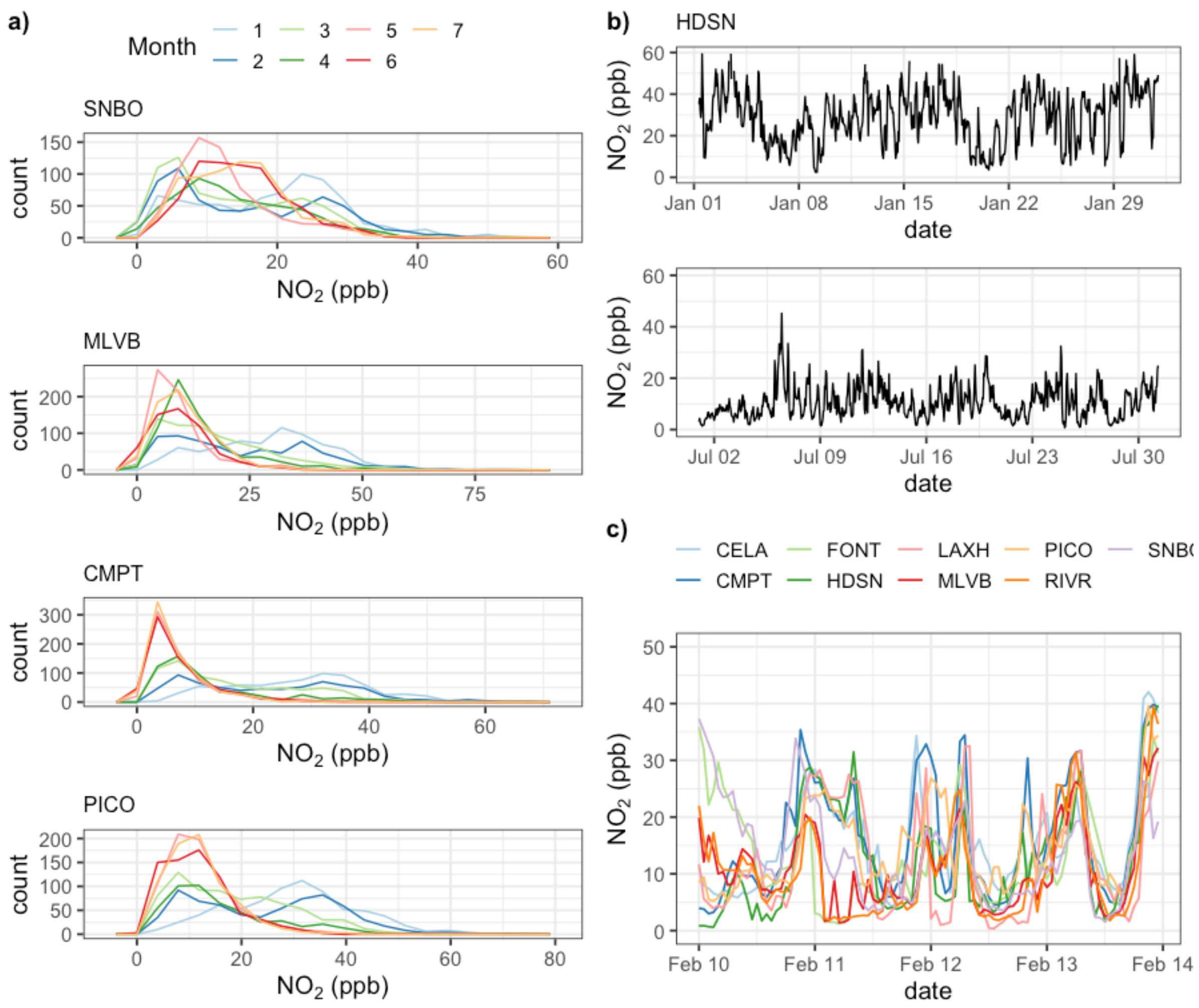

Figure 2. Examples of data for the spatio-temporal variation of $NO_2$. a) Frequency distributions of concentration at exemplar different sites month-by-month (month 1: January 2018; month 6: June 2018). b) Time series at the Hudson (HDSN) site, exemplifying variation in summer (January) and winter (July). (c) Time series over a few days at all nine sites, showing both similarities and differences across the study area.

*2.3 Proxy model*

A critical element of the framework is the proxy model, which is described in detail in Miskell et al., (2019). To summarise, if $X_{j,t}$ denotes the true concentration at site $j$ and time $t$, $Y_{j,t}$ denotes the sensor result, and $\hat{X}_{j,t}$ the estimate of $X$ derived from the measurement model, then the proxy model proposes that, over some time $t_d$ that is sufficiently long to average short-term fluctuations, a proxy site, $k$, can be identified, with data $Z_{k,t}$ such that the empirical cumulative probability distribution of $Z_k$ is a reliable estimate of the distribution of $X_j$, evaluated over $t_d$. Then the parameters of the measurement model can be estimated by adjusting them such that the distribution of $\hat{X}_j$ approximates the distribution of $Z_k$. In the previous work, the measurement model parameters were adjusted to match moments of $\hat{X}_{j,t}$ and $Z_{k.}$. Since we use only regulatory station data, where the instruments are frequently and rigorously calibrated, the simple measurement model applies:

$$\hat{X}_{j,t} = \hat{a}_0 + \hat{a}_1 Y_{j,t} + \varepsilon_{j,t} \quad (1)$$

with parameters obtained by matching the mean and variance of $\hat{X}_{j,t}$ and $Z_k$

$$\hat{a}_1 = \sqrt{\mathrm{var}\langle Z_{k,t-t_d:t}\rangle / \mathrm{var}\langle Y_{j,t-t_d:t}\rangle} \quad (2)$$

$$\hat{a}_0 = \mathrm{E}\langle Z_{k,t-t_d:t}\rangle - \hat{a}_1 \mathrm{E}\langle Y_{j,t-t_d:t}\rangle \quad (3)$$

Since $X$ is unknown, some means is required to identify proxies appropriate for measurement sites across the whole network, and to check their reliability given only the measurement results of the network. The simplest way is to compare results across the well-calibrated reference instruments using various choices of proxy for these. The regulatory network data are used to establish appropriate proxies for the low-cost network, which in turn would be used to extend the scope of the regulatory network to neighbourhood scale.

### *2.4 Proxy selection*

We explored two different approaches to select a proxy for $NO_2$. First, we used the closest proximity location regulatory site as a proxy. This approach has previously been used to correct $O_3$ data (Miskell et al., 2019; Miskell et al., 2018). However, given the high spatial variability of $NO_2$, the nearest regulatory site may not always be the most representative site. Thus, we also tested the applicability of proxies chosen based on land use similarities using the $k$-Nearest Neighbour classification ($k$NN). $K$NN is a supervised statistical learning technique aiming to classify the data to a given category based on a similarity using a test and a training set. The algorithm finds the $k$ training samples that are closest to the regulatory data of interest and assigns the most suitable proxy among the $k$ training examples to the regulatory site. Here, we have the land use data for each regulatory site (Table 1) and we use $k$NN to find a proxy for any regulatory site given its land use similarity. Here, we use $k$ = 2 (the second closest neighbour relative to a point is used). The land use data consisted of three variables: distance to motorway and freeway, primary road length within 1 km and elevation (Table 1). These variables were chosen based on a systematic literature review on land use regression (LUR) models developed for the North American Region. In total, significant covariates from 21 published $NO_2$ LUR studies (SI Table 1) were ranked to identify the most commonly reported land use variables explaining $NO_2$ variability in urban areas (Figure 3). As expected, the most commonly used predictors for $NO_2$ concentrations are related to traffic, length of major roads and distance to major road, followed by land use (commercial/industrial) and population density. The most used covariate was traffic, followed by major road length. However, local traffic (within 1 km) was not available for each site, thus we decided not to use traffic estimates for the proxy selection. Elevation improved the proxy selection and was included as well. Each variable was scaled so that they all had a similar range and were therefore comparable.

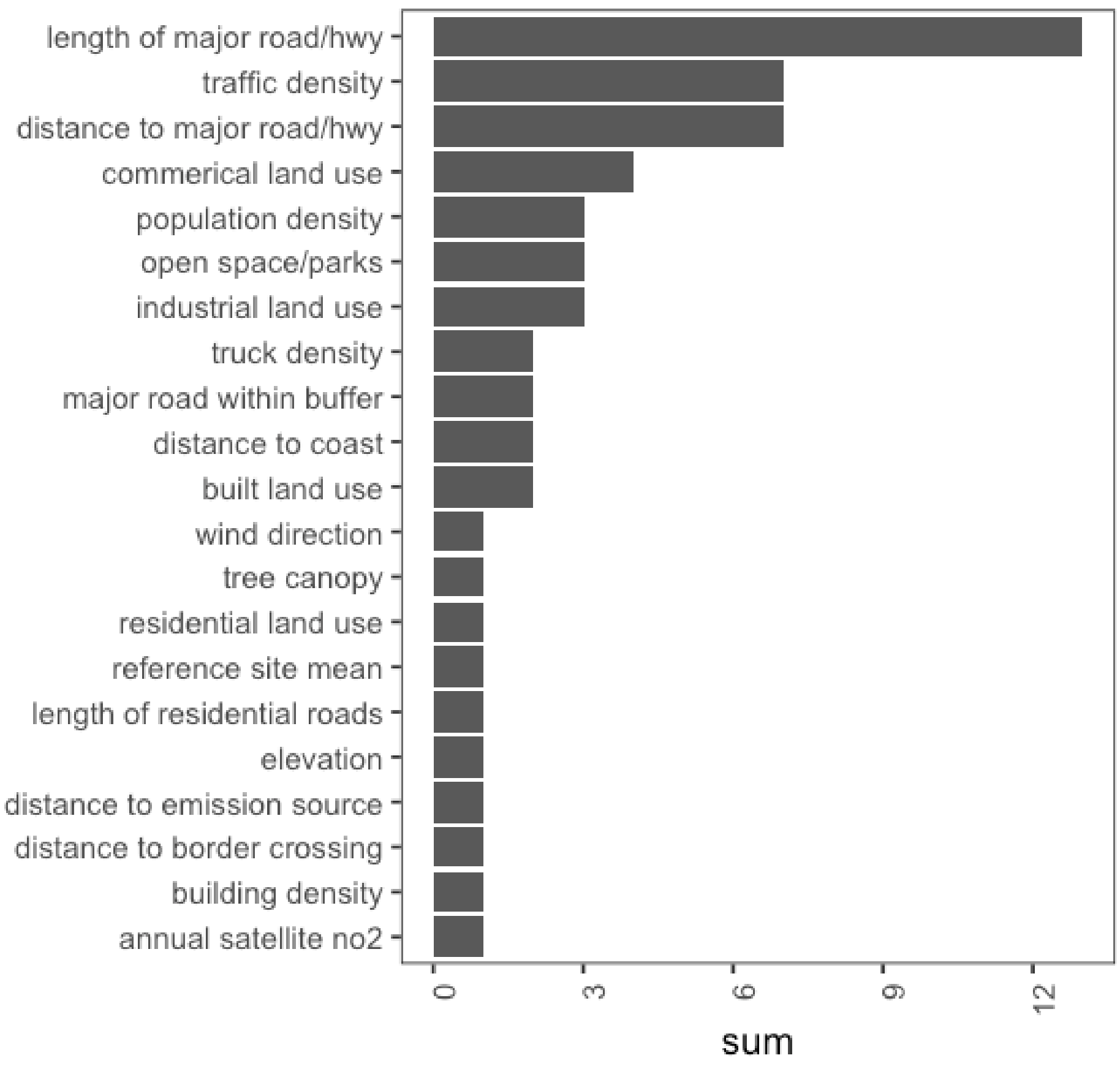

Figure 3. Commonly used covariates in LUR models to predict $NO_2$ concentrations in the North American region.

*2.5 Proxy validation*

We used two methods to evaluate the choice of proxy. First, we use the framework approach, which was introduced by Miskell et al. (2016) to detect drift. Given that regulatory data are used, the proxy signalling drift will in fact be an indicator of periods when the proxy site is not representative ("false alarm") and therefore this measure tests the performance of the different proxies. The approach uses simple statistical methods to compare the distribution (KS-test), and estimated slope and offset (mean and variance test: eq 2 and 3) between the site and its proxy, with an alarm raised if a defined threshold is exceeded. The thresholds were: $p^*_{KS}$ = 0.05, $\hat{a}_1$ = 1 ± 0.25, $\hat{a}_0$ = 0 ± 5 ppb. As a further test, we use the Kullback-Leibler divergence ($D_{KL}$) between the probability distribution of $X_j$ (the site of interest) and of $Z_k$, the proxy site being assessed, evaluated over the whole study period (January – July 2018). The distributions were constructed by computing normalised histograms of the data with defined bin width. The proxy site with the lowest $D_{KL}$ would have the best approximation of $Z_k$ to $X_j$ in the sense of maximum mutual information or minimum information entropy. If the proxy selection were appropriate, we would expect the selected proxy (selection based on land use variables or proximity) to also have the lowest $D_{KL}$. All statistical analysis was performed in R (v 3.5.2).

## *3. Results and Discussion*

The selected proxies for each regulatory site using the different selection methods are shown in Table 2. There is some overlap across the different proxy selection methods (e.g. PICO ≈ CELA, LAXH ≈ CMPT, CMPT ≈ HDSN, HDSN ≈ CMPT). The wide variation in $NO_2$ distribution across space and time, exemplified by the results given in Figure 2, illustrates the challenge in determining appropriate proxies for a region as varied as that of Southern California. However, Figure 4 shows the frequency distribution of the measured $NO_2$ concentrations over the 7-month period of the study at the regulatory sites compared to the frequency distribution of measured $NO_2$ concentrations at their proxy sites, and indicates that approaching the problem through a comparison of probability distributions over an appropriately chosen timescale indeed provides a way of defining suitable proxies. This is also supported by the $D_{KL}$, which was smallest between the site of interest and the proxy with the most similar land use (Table 2). An exception is LAXH, where the pollutant probability distribution was most similar to that measured at FONT (Table 2), suggesting that the land use similarity may not be representative of the pollutant distribution at this site. LAXH is the regulatory site at the Los Angeles International (LAX) airport and it is the least similar to any other regulatory sites in terms of land use and of pollutant distribution (ie largest KL divergence). The proxy site for SNBO is not as representative for lower $NO_2$ concentrations (< 25 ppb), however, these are also generally of less interest in air pollution studies.

Table 2. Proxy selection using different selection methods. Sites that are the same across different selection methods are highlighted in bold.

| **AQS ID** | **Nearest** | ***knn* (*k* = 2)** | **$D_{KL}$** |
|---|---|---|---|
| RIVR | MLVB | CELA | CELA |
| MLVB | RIVR | SNBO | SNBO |
| SNBO | RIVR | MLVB | MLVB |
| FONT | MLVB | SNBO | SNBO |
| PICO | **CELA** | **CELA** | **CELA** |
| CMPT | **HDSN** | **HDSN** | **HDSN** |
| LAXH | **CMPT** | **CMPT** | FONT |
| HDSN | **CMPT** | **CMPT** | **CMPT** |
| CELA | PICO | HDSN | HDSN |

The framework that we have proposed uses comparison over a rolling timescale of 3 days to signal an ‘alarm’ and 5 days to signal a ‘failure’. Thus Figure 5 is an overview of the number of times the framework approach signalled an alarm due to a threshold being exceeded (“false alarm”). These are “false alarms” because regulatory stations are being compared. Typically, the alarm was raised due to differences in the distribution of $NO_2$ concentrations between the site and its proxy (‘KS-test’), followed by a change in the slope (‘MV-slope’). As expected, the intercept (‘MV-intercept’) remained mostly stable between the regulatory site and its proxy site.

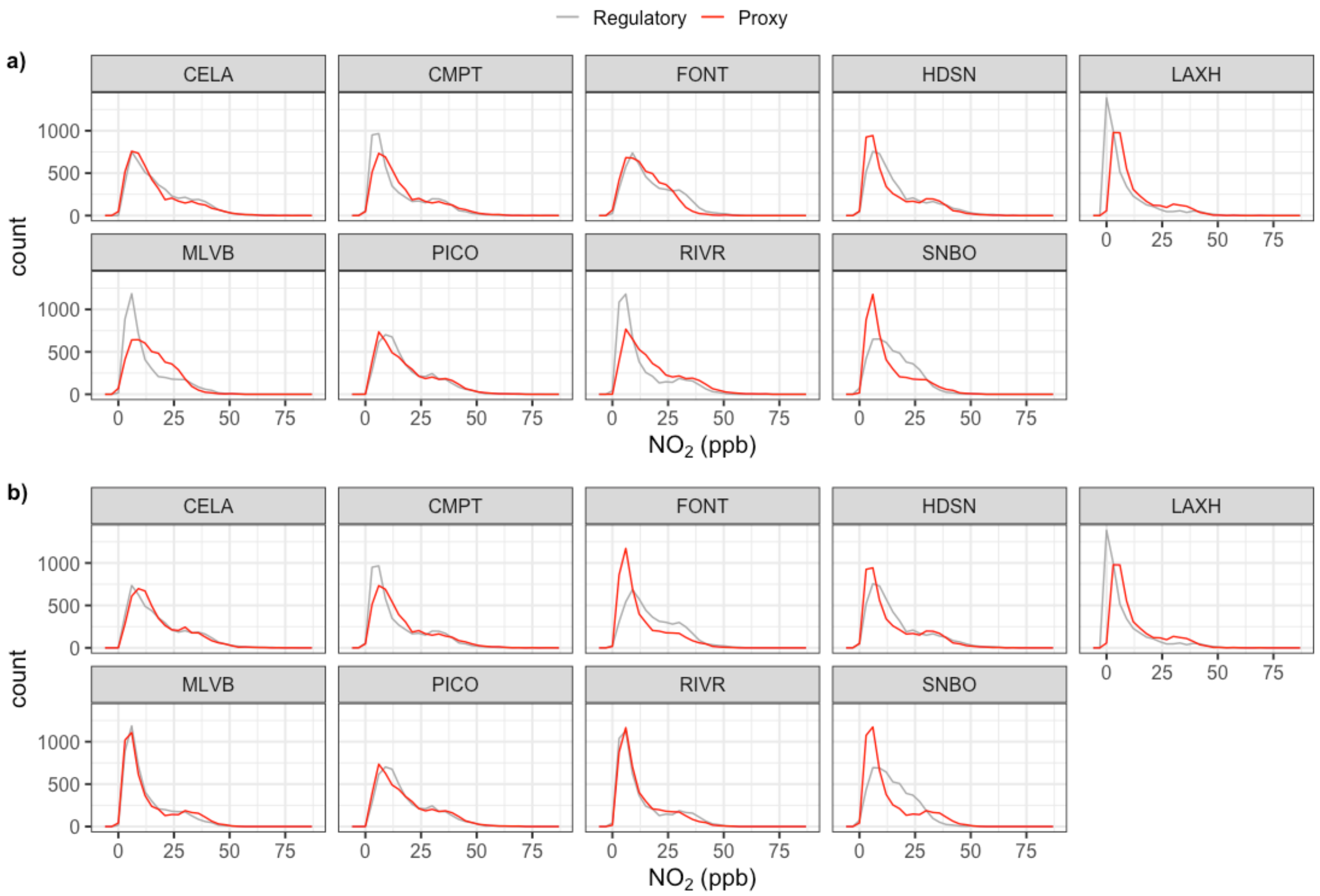

Figure 4. Frequency distribution of measured $NO_2$ concentrations at the regulatory sites and its proxy based on a) land use and, b) proximity.

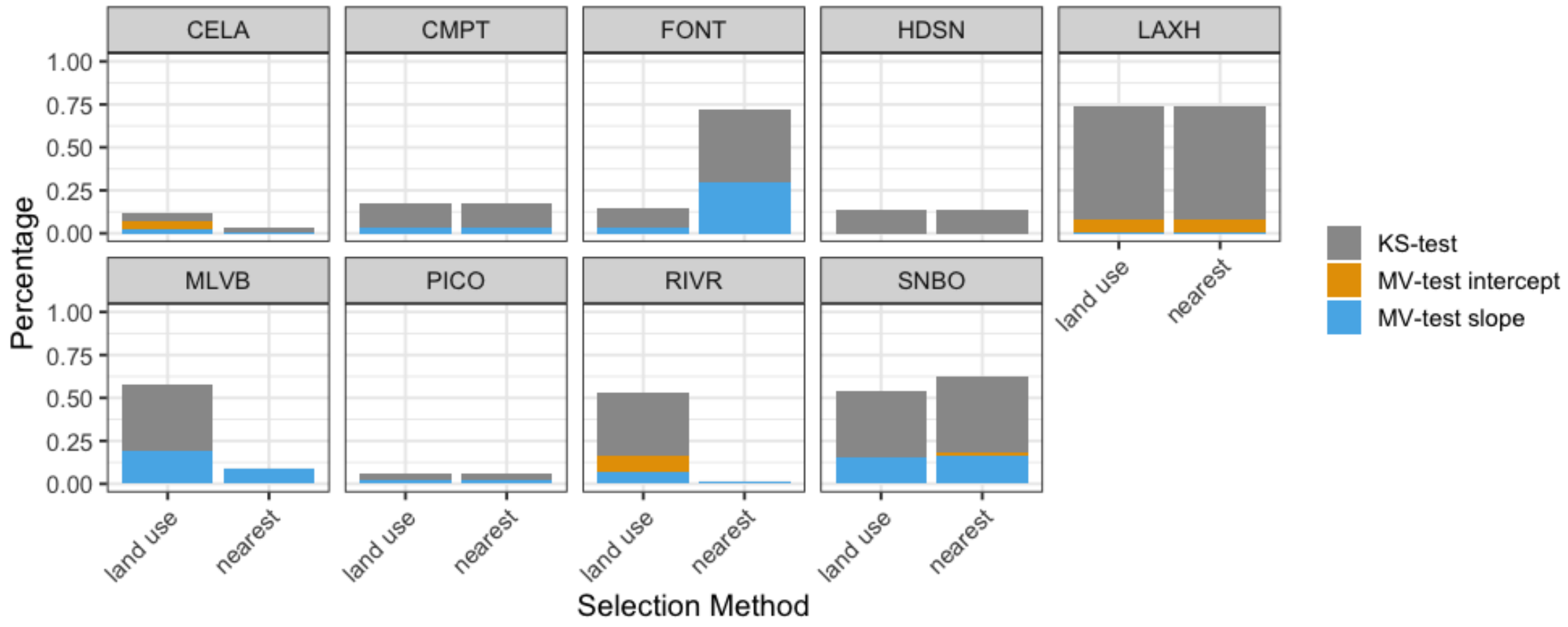

Figure 5. Bar chart showing how often the framework suggested drift ('false alarm') using the proxy data across the whole dataset (January – August 2018).

A better understanding of meteorological conditions associated with false alarms is provided when plotting the mean number of alarms signalled against wind direction and wind speed bins as shown in Figure 6. The proxy derived through $k$NN for RIVR, for example, is less suitable when wind was from the NE (Figure 5a): $NO_2$ at the site was significantly less than $NO_2$ at the proxy under these conditions. This is likely related to the mountains NE of the site and their effect on ozone transport and hence titration of vehicle-emitted NO. Further, false alarms were more common when wind speed was < 5 m $s^{-1}$ when observed $NO_2$ concentrations would be expected to be dominated by local emissions and possibly therefore different from those at a proxy site: again, possibly related to ozone transport. The relationship of false alarms to wind direction provides valuable insights that may be used to improve the proxy selection. For example, it may be possible to introduce conditional statements in the framework approach for situations when the proxy site is not suitable. For example, if land use NE of a site is different from its proxy site we may not want to correct any data when the wind direction is from NE. Since the land use choice for proxy agreed best with the smallest $D_{KL}$, we suggest using land use to select a proxy. An exception may be the sites situated in the valley surrounding the MLVB and RIVR (Mira Loma and Rubidoux region) regulatory sites where the nearest site was a more suitable proxy.

a)

b)

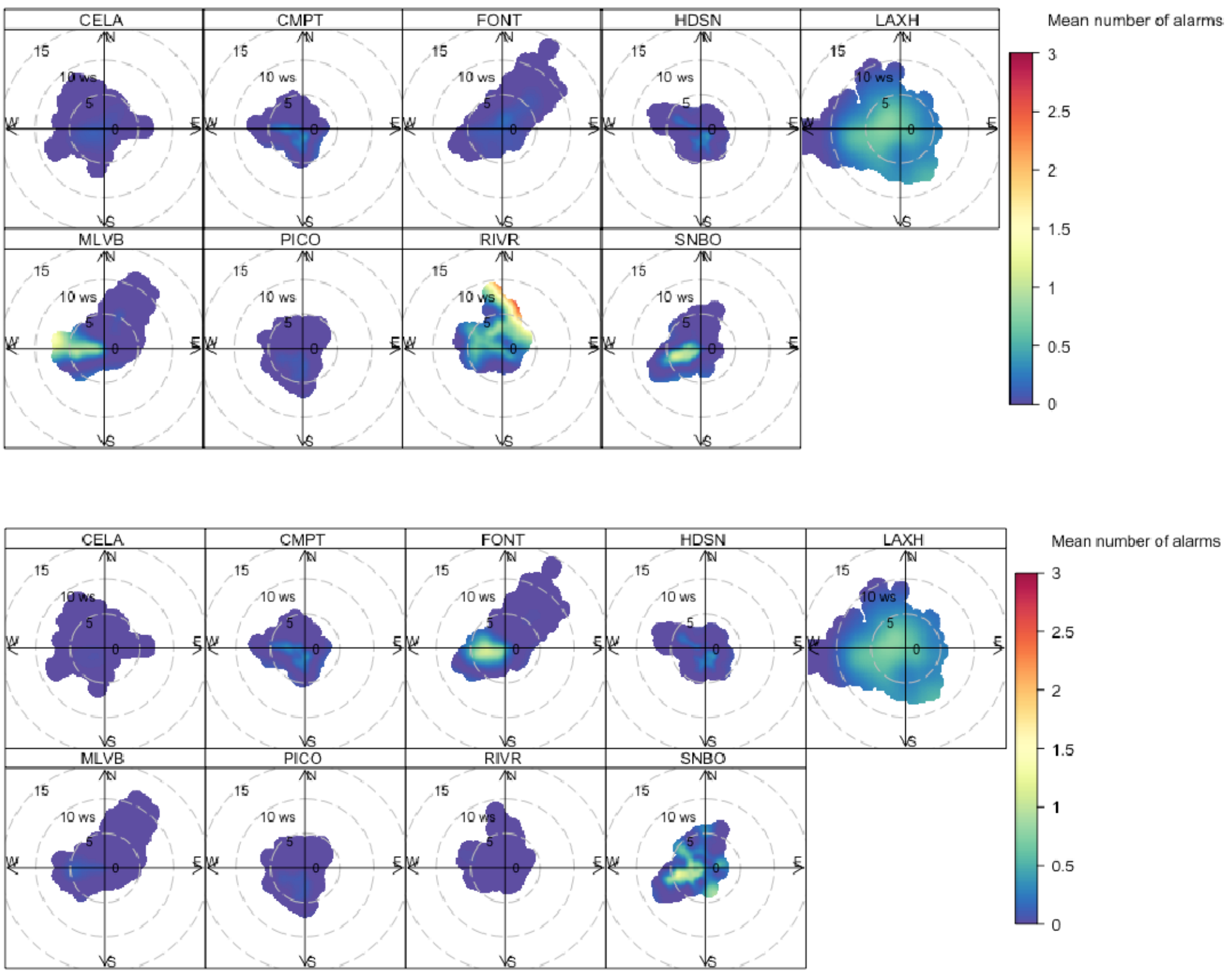

Figure 6. Polar plots showing the average alarm sum (from all three tests) by wind speed and wind direction across the whole measurement period. The mean is calculated for wind speed and direction bins. a) using the proxy with most similar land use, b) using the nearest proxy.

Figure 7 shows the framework corrected $NO_2$ concentrations at the regulatory sites using the proxy with the most similar land use (Figure 7a) and the closest proximity site as proxy (Figure 7b). We see that overall the framework approach worked well for most sites. If the proxy with similar land use is used for MLVB and RIVR, some $NO_2$ concentrations may wrongly be overestimated while at SNBO, low $NO_2$ concentrations (< 25 ppb) may be slightly underestimated. It is clearly visible that the slope between the framework corrected regulatory data and the original regulatory changes when the proxy fails (i.e. when the regulatory data is unnecessarily corrected). This suggests that the relationship between the variance at the regulatory site and the variance at the proxy site changed. As can be seen from Figure 7 the framework approach tended to fail during early summer (May/June/July) and when wind speed was low (< 3 m $s^{-1}$). At low wind speed the measured $NO_2$ concentrations will mostly be affected by local pollution sources, which may not be fully captured by the land use variables used in the *k*NN approach (distance to major road, road length within 1 km, elevation).

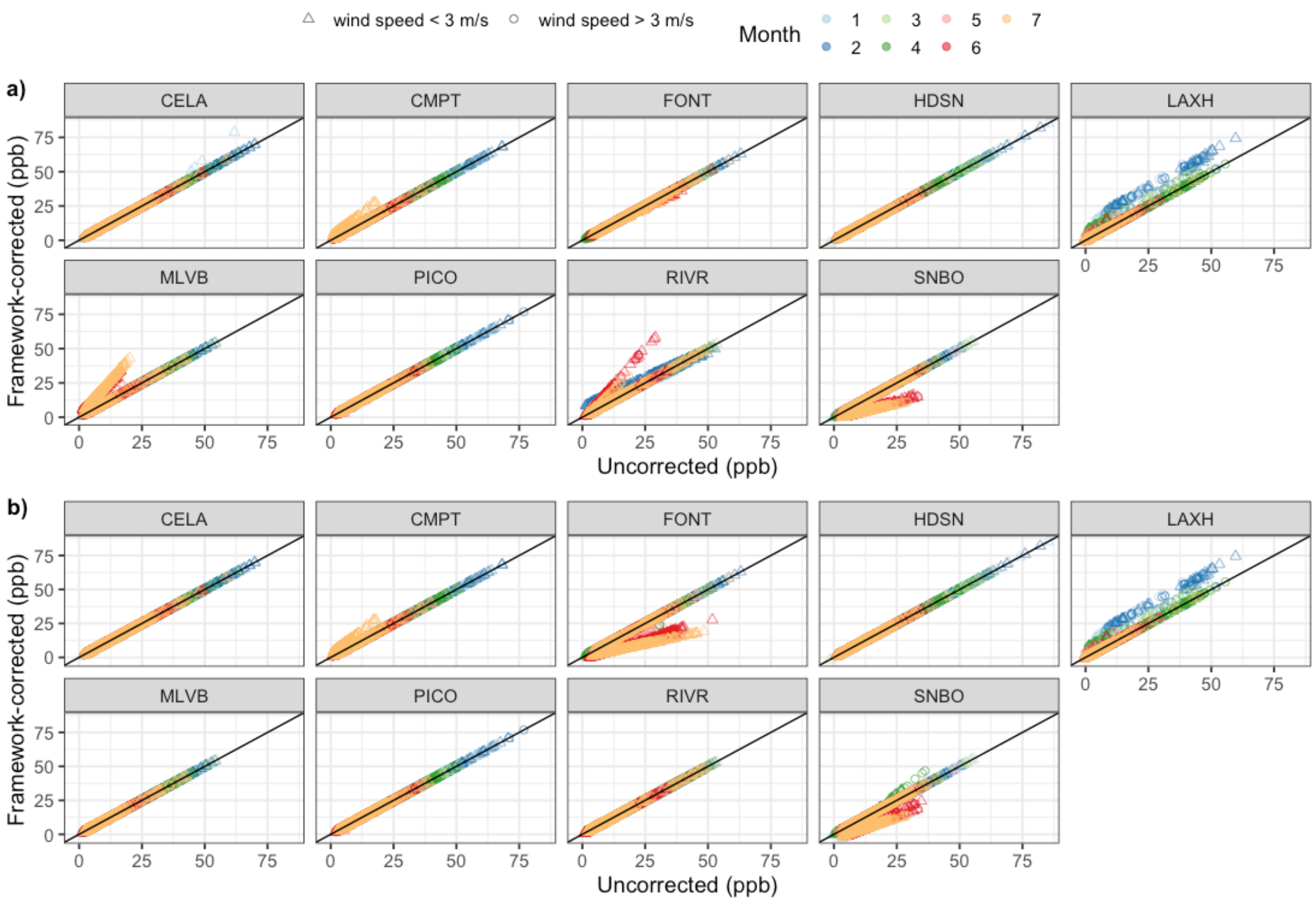

Figure 7 Scatterplots showing the framework corrected $NO_2$ concentrations (when > 1 test failed) against the regulatory concentrations coloured by month a) using the proxy with most similar land use, b) using the nearest proxy.

### *4. Conclusion*

This study has considered the problem of identification of suitable proxies for remote drift detection and calibration in a hierarchical network, proposed to comprise a few well-maintained regulatory instruments and a much denser network of low-cost sensors. The network of well-maintained regulatory instruments has been used to evaluate different methods of choosing proxies for a given site: specifically, proxy definition based on land use similarity, using covariates chosen by analysis of literature LUR studies of similar cities; and the simple choice of the closest proximity regulatory station to a given site as the proxy. The study has shown that, even in a region with high, and highly variable concentrations of air pollutants – nitrogen dioxide in Southern California - suitable proxies can be defined. Proxies based on land use similarity signalled typically less than 0.1% false alarms, except where the local geography was unusual – a semi-enclosed valley, for which the closest proximity station was an appropriate proxy – or when wind speed was low, when presumably local sources determined the concentration.

***Acknowledgments***

This work was funded by the New Zealand Ministry for Business, Innovation and Employment, contract UOAX1413. This work was performed in collaboration with the Air Quality Sensor Performance Evaluation Center (AQ-SPEC) at the South Coast Air Quality Management District (South Coast AQMD). The authors would like to acknowledge the work of the South Coast AQMD Atmospheric Measurements group of dedicated instrument specialists that operate, maintain, calibrate, and repair air monitoring instrumentation to produce regulatory-grade air monitoring data. DEW acknowledges the support of a fellowship at the Institute of Advanced Studies, Durham University, UK

**Supporting Information**

# Hierarchical network design for nitrogen dioxide measurement in urban environments, part 1: proxy selection

Lena Weissert[1,3], Georgia Miskell[1], Elaine Miles[2], Kyle Alberti[2], Brandon Feenstra[4], Hamesh Patel[3], , Vasileios Papapostolou[4], Andrea Polidori[4], Geoff S Henshaw[2], Jennifer A Salmond[3], David E Williams[1,*]

**Email* david.williams@auckland.ac.nz *ph +64 9 923 9877*

1. School of Chemical Sciences and MacDiarmid Institute for Advanced Materials and Nanotechnology, University of Auckland, Private Bag 92019, Auckland 1142, New Zealand

2. Aeroqual Ltd, 460 Rosebank Road, Avondale, Auckland 1026, New Zealand

3. School of Environment, University of Auckland, Private Bag 92019, Auckland 1142, New Zealand

4. South Coast Air Quality Management District, 21865 Copley Drive, Diamond Bar, CA 91765, USA

SI Table 1. LUR studies published for the North American region.

| First Author | Year Published | Study Location |
|---|---|---|
| (Gilbert et al., 2012) | 2005 | Montréal, Canada |
| (Sahsuvaroglu et al., 2012) | 2006 | Hamilton, Canada |
| (Jerrett et al., 2007) | 2007 | Toronto, Canada |
| (Henderson et al., 2007) | 2007 | Vancouver |
| (Mavko et al., 2008) | 2008 | Portland |
| (Wheeler et al., 2008) | 2008 | Windsor, Ontario |
| (Su et al., 2008) | 2008 | Vancouver |
| (Su et al., 2009) | 2009 | Los Angeles |
| (Mukerjee et al., 2009) | 2009 | Detroit/Dearborn |
| (Crouse et al., 2009) | 2009 | Montréal, Canada |
| (Wilton et al., 2010) | 2010 | Los Angeles |
| (Mercer et al., 2011) | 2011 | Los Angeles |
| (Novotny et al., 2011) | 2011 | United States |
| (Allen et al., 2011) | 2011 | Edmonton, Winnipeg |
| (Gonzales et al., 2012) | 2012 | El Paso, Texas |
| (Li et al., 2012) | 2012 | Southern California |
| (Clougherty et al., 2013) | 2013 | New York |
| (Beckerman et al., 2013) | 2013 | California |
| (Keller et al., 2015) | 2015 | Baltimore, Chicago, LA, NZ, St.Paul, Winston-Salem |
| (Deville Cavellin et al., 2016) | 2016 | Montréal, Canada |
| (Minet et al., 2017) | 2017 | Montréal, Canada |